\documentclass[epj]{svjour}
\usepackage{latexsym}
\usepackage{graphics}
\begin{document}
\title{Superconducting-Insulator Transition in Disordered Josephson Junctions Networks}
\titlerunning{Superconducting-Insulator Transition in Disordered Josephson Junctions Networks}
%Insert here a short version of the title if it exceeds 70 characters
\author{L. Ponta \inst{1} \and V. Andreoli \inst{2} \and A. Carbone \inst{1,3,4}}
\institute{\inst{1} Physics Department, Politecnico di Torino, Corso Duca degli Abruzzi 24, 10129 Torino, Italy\\
\inst{2} Istituto Nazionale di Ricerca Metrologica (INRIM), Strada delle Cacce 91, 10135 Torino, Italy \\
\inst{3} Istituto dei Sistemi Complessi, ISC-CNR Roma, Italy\\
\inst{4} ETH Zurich, Switzerland}
\authorrunning{Ponta, Andreoli, Carbone}
\abstract{The superconducting-insulator transition is simulated in disordered networks of Josephson junctions with thermally activated Arrhenius-like resistive shunt. By solving the conductance matrix of the network, the transition is reproduced in different experimental conditions by tuning thickness, charge density and disorder degree.
In particular,  on increasing fluctuations of the parameters entering the Josephson coupling and the Coulomb energy of the junctions, the transition occurs for decreasing values of the critical temperature $T_c$ and increasing values of the activation temperature $T_o$.  The results of the simulation compare well with recent experiments where the mesoscopic fluctuations of the phase have been suggested as the mechanism underlying the phenomenon of \emph{emergent granularity} in otherwise homogeneous films. The proposed approach is compared with the results obtained on TiN films and nanopatterned arrays of weak-links, where the superconductor-insulator transition is directly stimulated.}
\maketitle
Although superconductivity has been predicted to persist even in the presence of impurities \cite{Anderson}, a superconducting insulating transition (SIT) has been observed in strongly disordered metals \cite{Finkelstein,Jaeger,Fisher,Trivedi,Wallin,Yazdani,Hsu}.
A `fermionic' phenomenon explaining the superconducting-insulating transition is related to the suppression of the Cooper pairing by the enhancement
of the Coulomb repulsion between electrons with increasing disorder. An alternative `bosonic' phenomenon driving the
transition is related to the phase fluctuations that could cause pair breaking and destroy the superconducting energy gap.
Spatial inhomogeneities of the order parameter might lead to complex effects due to the formation of islands with nonnegligible superconducting order parameter embedded in an insulating matrix \cite{Ghosal,Galitski,Skvortsov,Sacepe,Dubi,Bouadim}.
\par
Recent investigations carried out in homogeneous thin films by tuning disorder, electric or magnetic field  pointed to a mechanism of \emph{self-induced granularity} \cite{Shahar,Sambandamurthy,Baturina,Baturina2,Sarwa}. The emergent collective insulating state exhibits threshold voltage depinning behavior and thermally activated resistance
described by an \emph{Arrhenius-like exponential} law.
In spite of the homogeneity, such systems have been thus envisioned to behave as networks of Josephson junctions, where the onset of the insulating phase requires the Coulomb energy $E_C$:
\begin{equation}
\label{Ec}
E_C = \frac{e^2}{2C}
\end{equation}
be larger than the coupling energy of the Josephson junction:
\begin{equation}
\label{EJ}
E_J = \frac{\pi}{4} \left(  \frac{h}{e^2 R_n}\right)\Delta \hspace{5pt}.
\end{equation}
Different mechanisms leading to the superconducting-insulating transition very likely coexist in such complex electronic systems with collective insulating states emerging within the homogeneous structure.
\par
\begin{figure}
\begin{center}
\resizebox{0.75\columnwidth}{!}
{
\includegraphics{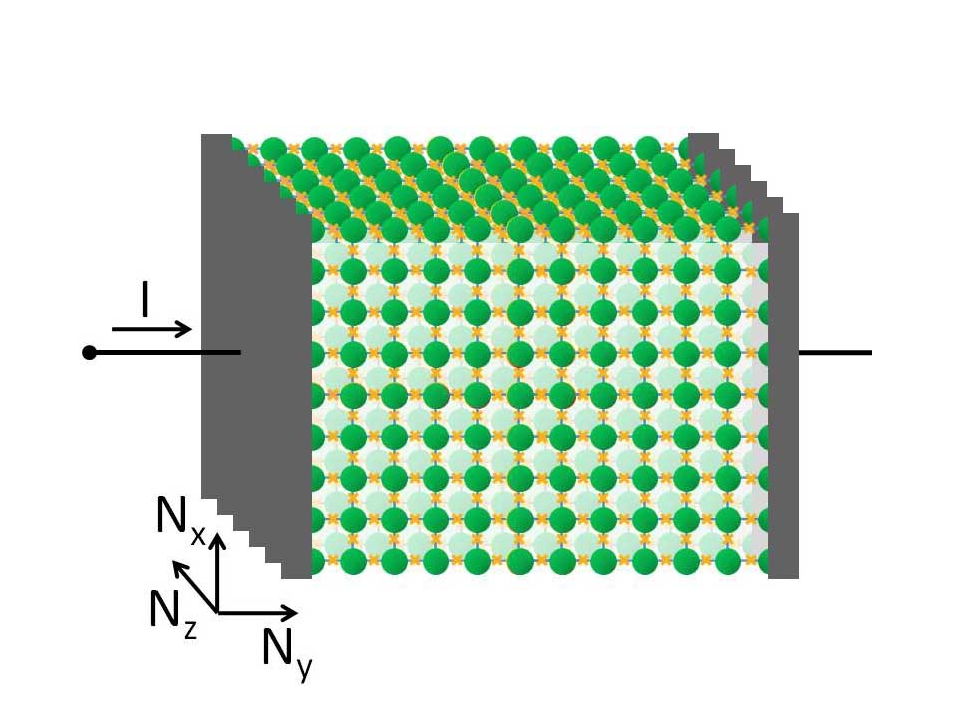}
}
\caption{Scheme of a three-dimensional network of Josephson junctions. Circles are the superconducting nodes. The links between nodes behave as Josephson junctions. $N_{x}$, $N_{y}$ and $N_{z}$ indicate the number of Josephson junctions respectively in the $x$, $y$ and $z$ directions. The network is biased by a constant current source with intensity  $I$.}
\label{network}
\end{center}
\end{figure}
\begin{figure}
\begin{center}
\resizebox{0.75\columnwidth}{!}
{
\includegraphics{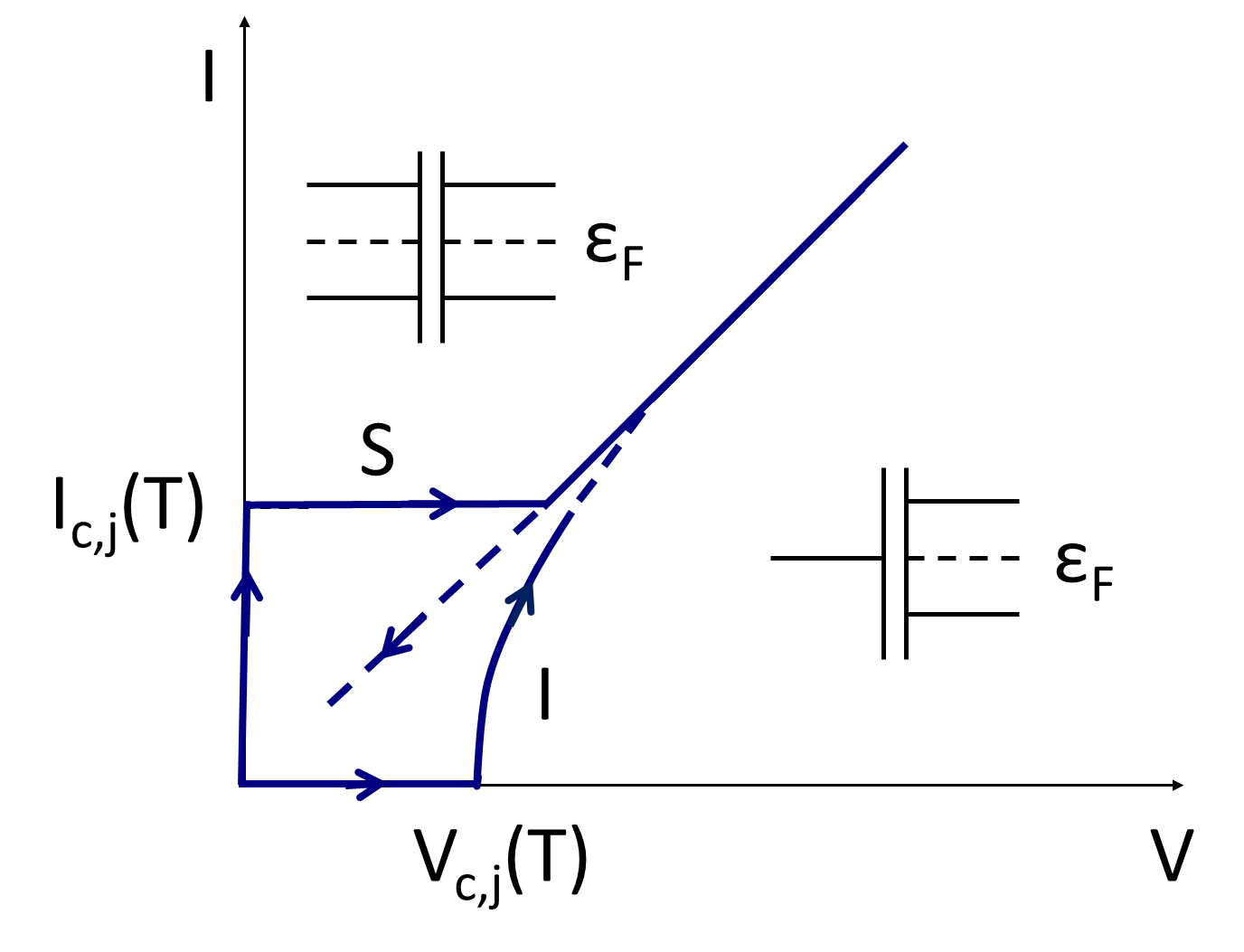}
}
\caption{Scheme of the current-voltage characteristic of a Josephson junction for the
superconductor-normal (upper branch) and insulator-normal (lower branch) states. The dashed line is obtained when the external drive decreases as an effect of the  hysteresis. Due to the disorder of the  network, a wide variability of the I-V characteristics should be assumed according to the probability distribution function of the  relevant parameters of the Josephson junctions. In the present work, a Gaussian probability distribution has been considered.}
\label{IV}
\end{center}
\end{figure}

In this work,  the ``self-induced granularity" is modelled by disordered networks of Josephson junctions \cite{Efetov,Bradley,Granato,Yu,Otterlo,vanDerZant,Beloborodov,Syzranov}.
The superconductor-insulator transition is simulated by solving the conductance matrix of Josephson junctions.
The networks are biased by a constant current generator $I$, as sketched in Fig.~\ref{network}.
The superconductive, intermediate and resistive state of each Josephson junction is set according to each single current-voltage characteristic which is shown in Fig.~\ref{IV}. The approach was previously used to investigate the normal-superconductive  transition and the current fluctuation power spectra in \cite{Ponta}.
\par
It is shown that an Arrhenius-like dissipation mechanism in the shunted Josephson junction induces the insulating phase.
Furthermore, the resistance variation observed in the array well compares with the experimental results of the  superconducting-insulator transition
in homogeneous and nanopatterned TiN films \cite{Baturina,Baturina2}.
As already mentioned, the resistance in the insulating state depends on temperature according to an \emph{Arrhenius-like exponential} as found in many experiments \cite{Shahar,Sambandamurthy,Baturina,Baturina2,Sarwa}. The resistance $R$ of the Josephson junction array is proportional to the resistance $R_{j}$  of each weakly linked Josephson junction according to the relationship:
\begin{equation}
\label{Rj}
R=R_j\frac {N_x N_z+1}{N_y}\hspace{5pt},
\end{equation}
where the subscript $j$ identifies a single junction, while $N_{x}$, $N_{y}$ and $N_{z}$ indicate the total number of Josephson junctions respectively in the $x$, $y$ and $z$ directions  as shown in Fig.~\ref{network}. The resistance $R_j$ of each Josephson junction is taken of the form:
\begin{equation}
\label{eq1}
R_j = R_{oj}\exp\left(\frac{T_{oj}}{T}\right)^{\gamma}  \hspace{5pt},
\end{equation}
with $T_{oj}$  the activation temperature and $R_{oj}$ the normal state resistance of each junction. The exponent $\gamma$ takes values in the range $0$ to $1$. Specifically, $\gamma=1$ corresponds to the Arrhenius law, while fractional values of $\gamma$ feature variable range hopping transport. Experiments have shown that $\gamma=1$ holds
on the insulating side of the transition at low temperatures, while variable range hopping transport with exponent $\gamma=1/2$ or  $\gamma=1/4$
are likely to occur over the normal side of the transition at higher temperatures. The activation temperature $T_{oj}$ is related to the parameters of the junction through:
\begin{equation}
\label{To}
T_{oj} \sim \Delta_j+ E_{Cj}/4  \hspace{5pt}.
\end{equation}
This relationship means that a Cooper pair, in the superconducting islands separated by insulating barrier of  each junction $j$, is firstly broken by supplying an energy $\Delta_j$.  Then the charge is redistributed in each tunnel capacitance ${C_j}$ by supplying an additional amount of energy $E_{Cj}/4$.  When the Coulomb energy is larger than the Josephson coupling, the value of the activation temperature is thus mostly determined by the value of the junction capacitance.
\par
It is worth noting that the dimensionless tunneling conductance $g$ of the  Josephson junction is defined as:
\begin{equation}
\label{g}
g = \frac{G}{e^{2}/h}  \hspace{5pt},
\end{equation}
\noindent
where $G$ is the average conductance  between adjacent grains and $e^{2}/h$ is the quantum conductance. The average conductance $G$ of the array can be obtained as the reciprocal of average resistance $R$ defined by Eq.~(\ref{eq1}).
\par
It has been observed both theoretically and experimentally that when the average tunneling conductance $G$ is greater than the quantum conductance, i.e.  when $g\gg1$, which corresponds to strong coupling between grains, a superconducting state is achieved at low temperature. Conversely, when $g\ll 1$, which corresponds to weak coupling between grains, an insulating state emerges at low temperature \cite{Beloborodov}.
This phenomenon can be explained by the existence of an additional dissipative channel due to single electrons tunneling between grains,  resulting in an  \emph{effective Coulomb energy}:
\begin{equation}
\label{ECeff}
\widetilde{E}_{C} = \frac{\Delta}{2g}  \hspace{5pt}.
\end{equation}
By comparing Eqs. (\ref{EJ}) and (\ref{ECeff}),   one can note that $E_J$ is always larger than $\widetilde{E}_{C}$ for $g\gg1$, implying a superconducting ground state.  Conversely, for $g\ll1$, $E_J$ is smaller than $\widetilde{E}_{C}$, allowing the onset of an insulating state.
Both the numerical and qualitative description clearly show, that a superconducting insulator transition occurs as $E_J$ compares with $k_BT$ (disorder and temperature induced superconducting insulator transition) which ultimately corresponds to compare Josephson and Coulomb coupling through the average tunneling conductance $G$ of the array and the consequent coexistence of the fundamental mechanisms underlying the superconducting insulator transition described above.
\par
The state (superconductive, normal, insulating) of each Josephson junction in the array is set according to each  current voltage (I-V) characteristics. In the I-V curve represented in Fig.~\ref{IV},  $I_{c,j}$  and $V_{c,j}=I_{c,j}R_j$  represent the critical current and the critical voltage of a single Josephson junction $j$.
For the superconductive branch (upper branch)  of the I-V characteristics, the normal resistive state is achieved when the bias current $I_j$ flowing through the junction $j$ exceeds the critical current $I_{c,j}$. For the insulating branch (lower branch) of the I-V characteristics,  the normal resistive state is achieved when the bias voltage $V_j$ across each junction $j$
exceeds the critical voltage $V_{c,j}$.
\par
The critical current is a function of temperature according to the following linearized equation:
\begin{equation}
\label{Ic_critical}
I_{c,j}=I_{co,j} \left(1- \frac{T}{T_c} \right) \hspace{5pt},
\end{equation}
\noindent
where $I_{co,j}$ is the lowest temperature critical current. An analogous relationship holds for the critical field $H_{c,j}$.
To simulate the network disorder,  the critical
current $I_{c,j}$, the critical field $H_{c,j}$ and the activation temperature $T_{o,j}$  are taken as random variables, which fluctuate around their average value
according to a Gaussian distribution with standard deviation $\sigma$ \cite{Ponta}.
\par
The average resistance $R$ of the network depends on elementary shunt resistance $R_{j}$ through the linear relationship Eq.~(\ref{Rj}). Therefore, the variation $\Delta R$, which is a measure of the slope of the transition, depends on
bias current $I$, on temperature $T$ and magnetic field $H$, through the variation of the elementary shunt resistance $\Delta R_{j}$ of each weak-link:
\begin{equation}\label{eq3}
\Delta R_{j}= \frac{\partial R_{j}}{\partial I} \Delta I+ \frac{\partial R_{j}}{\partial H} \Delta H + \frac{\partial R_{j}}{\partial T} \Delta T \hspace{5pt}.
\end{equation}
\noindent The terms on the right hand side of Eq.~(\ref{eq3}) will be now related to the relevant parameters of the junctions  $I_{c,j}$ and $H_{c,j}$ used in the simulation. The first and second terms on the right hand side of Eq.~(\ref{eq3}) can be written as $(\partial R_{j} / \partial I_c) \Delta I_c$ and $ (\partial R_{j} / \partial H_c) \Delta H_c$.  The term $\partial R_{j}/\partial T$ can be written as derivative of a compound function:
\begin{equation}
\frac{\partial R_{j}}{\partial T} = \frac{\partial R_{j}} {\partial I_c}  \frac{\partial I_c} {\partial T} + \frac{\partial R_{j}}{ \partial H_c}  \frac{\partial H_c}{ \partial T}\hspace{5pt}.
\end{equation}
Moreover, from Eq.~(\ref{Ic_critical}),  $\partial I_c/\partial T \simeq -I_{co}/T_c$. A similar relation holds for the derivative of the critical field with respect to $T$.
On account of the relation $\Delta I_c = I_c - I_{co} \propto \sigma$ and $\Delta H_c = H_c - H_{co} \propto \sigma$, one obtains that the slope of the resistive transition $\Delta R$  is related to the disorder degree $\sigma$.
Since it is $(\partial R_{j} / \partial I_c) < 0$ and $(\partial R_{j} / \partial H_c) < 0$, Eq.~(\ref{eq3}) shows that if either $\Delta I_c$  or $\Delta H_c$ increases, $\Delta R_{j}$ decreases.
\par
The quantity $\Delta R_{j}$ turns out to be proportional to the slope of the resistive transition curve, which is smoother when either $\Delta I_c$ or $\Delta H_c$ increases, i.e.  when the disorder increases.  Therefore, the disorder  enters the simulation directly by varying the parameter $\sigma$.
The disorder degree is strongly dependent upon the temperature $T$ along the transition as the Josephson phase fluctuations depend on $T$. For the sake of simplicity and because the work is mostly aimed at reproducing the transition from the superconducting to the insulating phase at a given low temperature,  the simulations have been performed at constant value of $\sigma$ along the superconducting normal transition curve.
\par
 Additionally, the SIT transition is shown to occur either when thickness decreases  or when current increases, while the parameter $\sigma$ is kept constant. The aim of these simulations is to show that external drives, such as bias current or thickness variation, directly act on the network as ``disorder enhancer" and  trigger the onset of the insulating phase.  The temperature dependence of the disorder ($\sigma$)  would further accelerate the occurrence of the insulating phase in such cases as well.
%
%
%%%%%%%%%%%%%%%%%%%%%%%%%%%%%%%%%%%%%%%%%%%%%%
%%%%%%%%% SECTION III.  RESULTS %%%%%%%%%%%%%%
%%%%%%%%%%%%%%%%%%%%%%%%%%%%%%%%%%%%%%%%%%%%%%
%
%
\par
The simulations of the superconductor-insulator transition in the Josephson junction arrays are performed as a function of the temperature $T$ in the range $0$ to  $1 \mathrm{K}$ by using different tuning parameters, as electric field, thickness and disorder degree. The resistive shunt is taken of the standard Arrhenius law form with $\gamma=1$.
\par
\begin{figure}
\begin{center}
\resizebox{0.95\columnwidth}{!}{%
\includegraphics{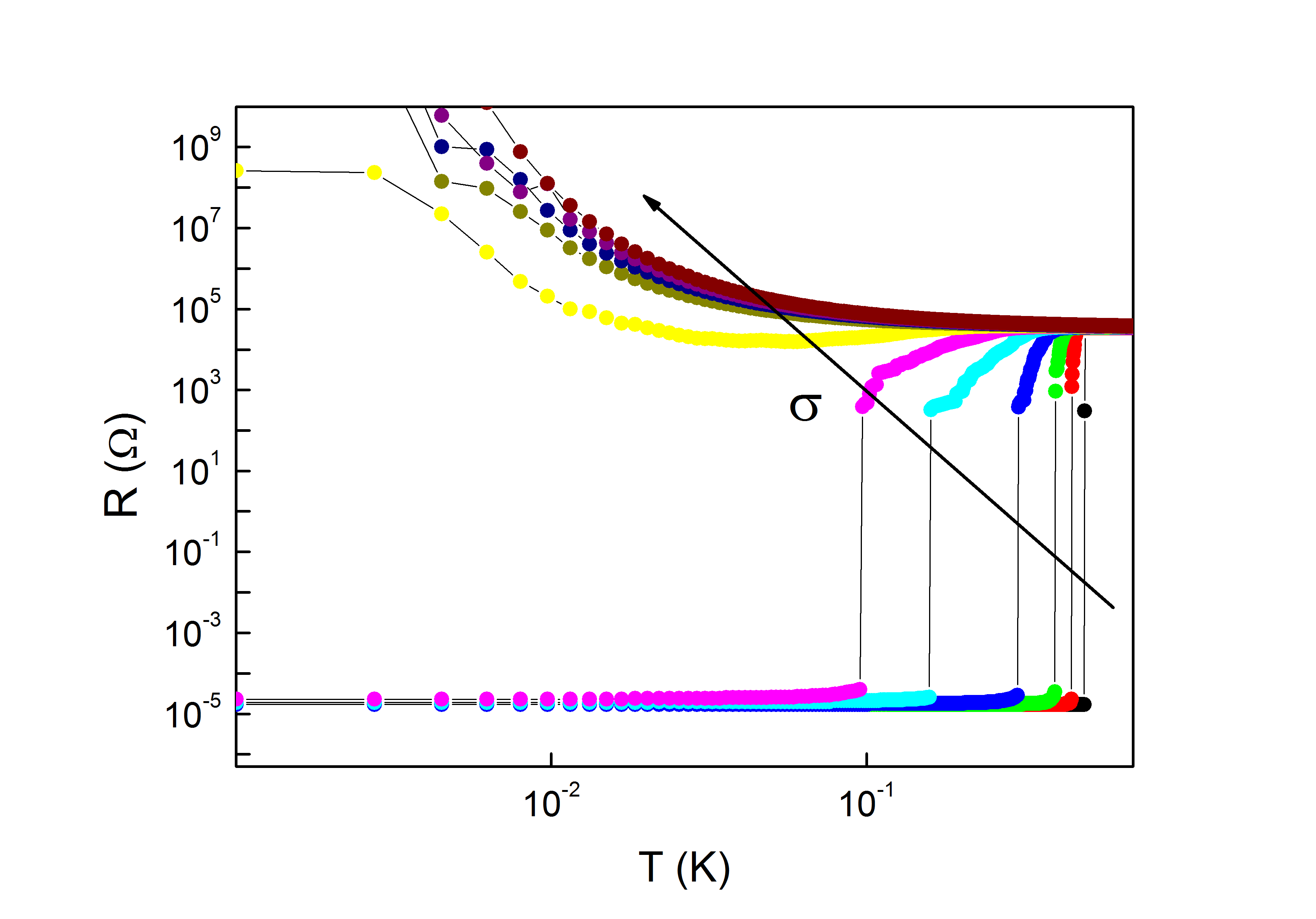}
}
\caption{Superconducting-insulator transition of a two-dimensional network of Josephson junctions as a function of temperature. For all the curves, it is  $N_{x}=30$, $N_{y}=50$, $N_{z}=1$. The bias current is kept constant $I=9.5\mu\mathrm{A}$. Different curves correspond to
different degrees of disorder with the standard deviation of the critical currents $\sigma$ varying from 0 to 1 with step $0.1$.   }
\label{PhaseTransitionSigma}
\end{center}
\end{figure}

Figure~\ref{PhaseTransitionSigma} shows the superconductor-insulator transition when the temperature increases and the bias current $I$ is kept constant in two-dimensional disordered networks. Different curves correspond to different disorder, i.e. to different values of $\sigma$.
By increasing temperature, the current $I_{c,j}$ of each junction decreases according to Eq.~(\ref{Ic_critical}).  As $I_{c,j}$ becomes smaller than the bias current flowing through each junction, the junction becomes resistive.
In the ideal case of a network with no disorder, it is $\sigma=0$ and the transition curve is vertical. All the Josephson junctions have the same critical current, become resistive at the same temperature and the transition occurs simultaneously all through the network. For the curves shown in Fig.~\ref{PhaseTransitionSigma}, the average value of the normal resistance $R_o$ is taken equal to $20 \mathrm{k \Omega}$ and the average value of the critical temperature  $T_c =0.8$ for $\sigma=0$. Moreover, when $\sigma=0$  the activated conduction processes are negligible, so that the average value of the activation temperature can be taken equal to zero, $T_o\simeq 0$. The average value of the critical current $I_{c,o}$ is $20 \mu A$.
\par
As disorder increases ($\sigma$ increases accordingly), the resistance changes more smoothly due to the wider spread of the $I_{c,j}$ values. Upon further disorder increase, the transition becomes smoother and smoother and for even larger disorder the network becomes an insulator.
As observed in many experiments, in the case of the standard Arrhenius law, the average value of the activation temperature $T_o$ is of the same order of magnitude of $T_c$. Both $T_c$ and $T_o$ are related to the disorder in the film, however $T_c$ decreases while the average value of $T_o$ increases as disorder increases. One can assume $T_o \propto \sigma$ implying that $T_o$ increases when disorder increases.  By increasing $\sigma$ the value of the critical temperature $T_c$ decreases accordingly. The average values of the activation temperature $T_o$ corresponding to each curve range from $0$, for $\sigma=0$, to $8$, for $\sigma=1$, with step 1.
The values of the parameters used in these simulations are in the range of experimental values reported in \cite{Baturina,Baturina2}.
%
%
% Figura 3
%
%
\par
\begin{figure}
\begin{center}
\resizebox{0.95\columnwidth}{!}{%
 \includegraphics{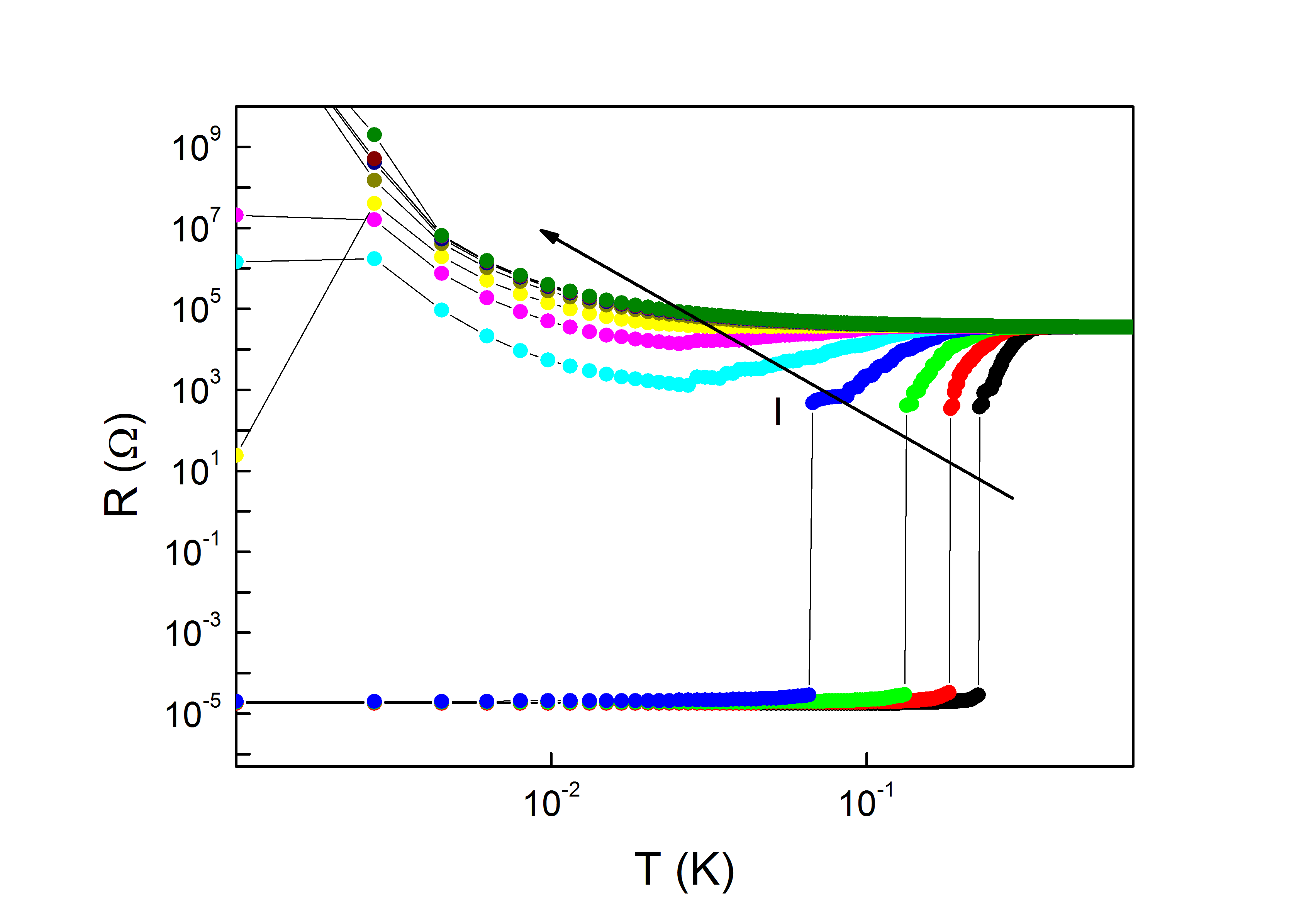}
}
\caption{Superconducting insulator transition of a two-dimensional network of Josephson junctions as a function of temperature. For all the curves, it is  $N_{x}=30$, $N_{y}=50$, $N_{z}=1$.  The degree of disorder is kept
constant by taking the standard deviation of the critical currents $\sigma$ equal to 0.3. Different curves refer to
different values of the bias current $I$ ranging from  $11\mu\mathrm{A}$ to $22\mu\mathrm{A}$ with steps $1\mu\mathrm{A}$. }
\label{PhaseTransitionIb}
\end{center}
\end{figure}

Figure~\ref{PhaseTransitionIb} shows the superconducting insulator transition as a function of the temperature, for different values of the bias current $I$.
The degree of disorder is kept constant by taking constant the value of $\sigma$ (in particular $\sigma= 0.3$). By increasing the bias current, the resistive transition becomes smoother
and the superconductor insulator transition occurs at larger values of the bias current.
%
%
% Figure 4
\par
\begin{figure}
\begin{center}
\resizebox{0.95\columnwidth}{!}{%
\includegraphics{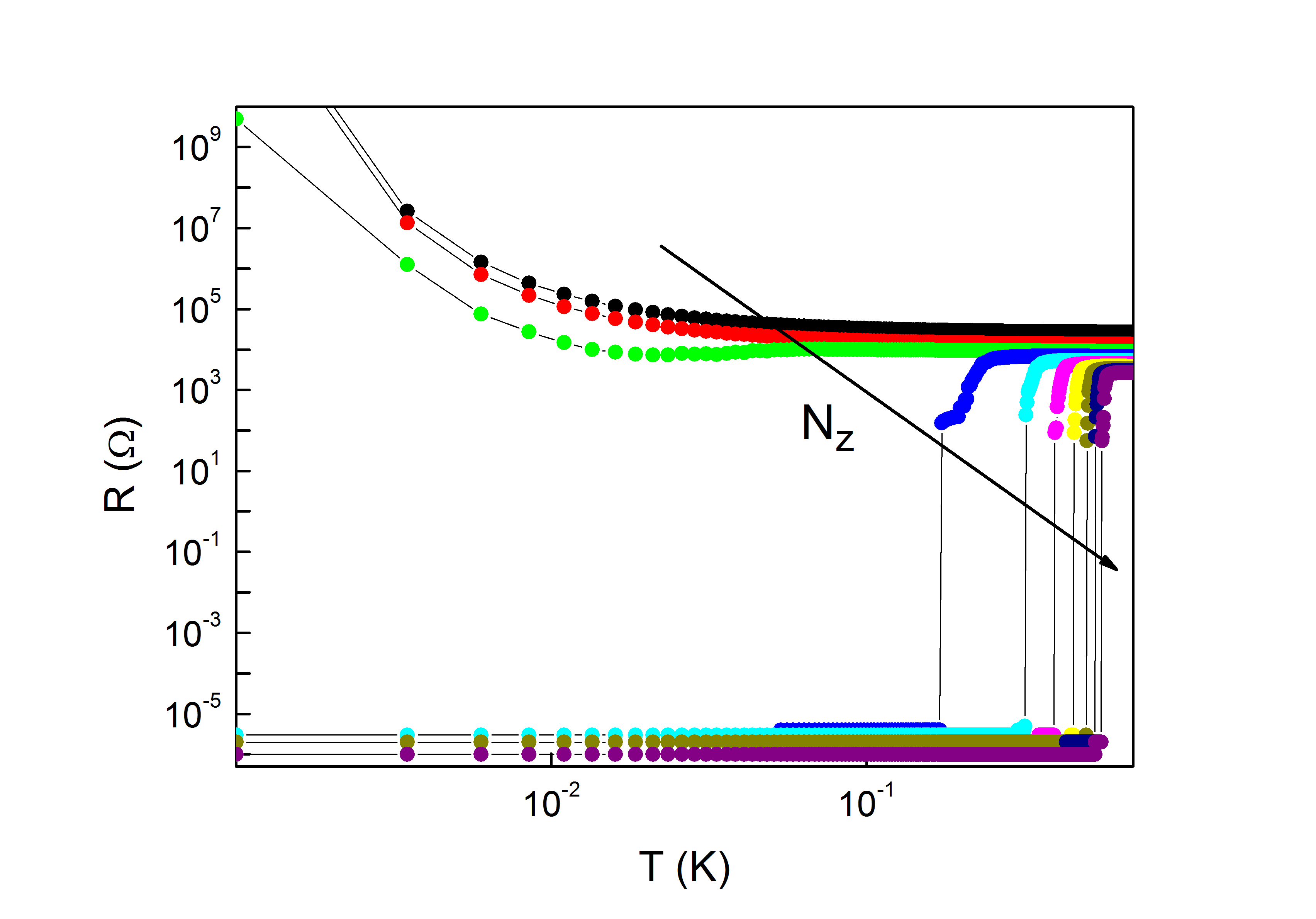}
}
\caption{Superconducting-insulator transition of a three-dimensional network as a function of temperature. For the curves shown in this figure,  $N_{x}=20$, $N_{y}= 25$  whereas $N_z$ is varied from 1 to 10 in steps of 1 to simulate the film thickness variation. The bias current is kept constant to the value $I=50\mu\mathrm{A}$. The degree of disorder is kept constant with
the standard deviation $\sigma=0.3$.}
\label{PhaseTransitionNzSigma03}
\end{center}
\end{figure}
\begin{figure}
\begin{center}
\resizebox{0.95\columnwidth}{!}{%
\includegraphics{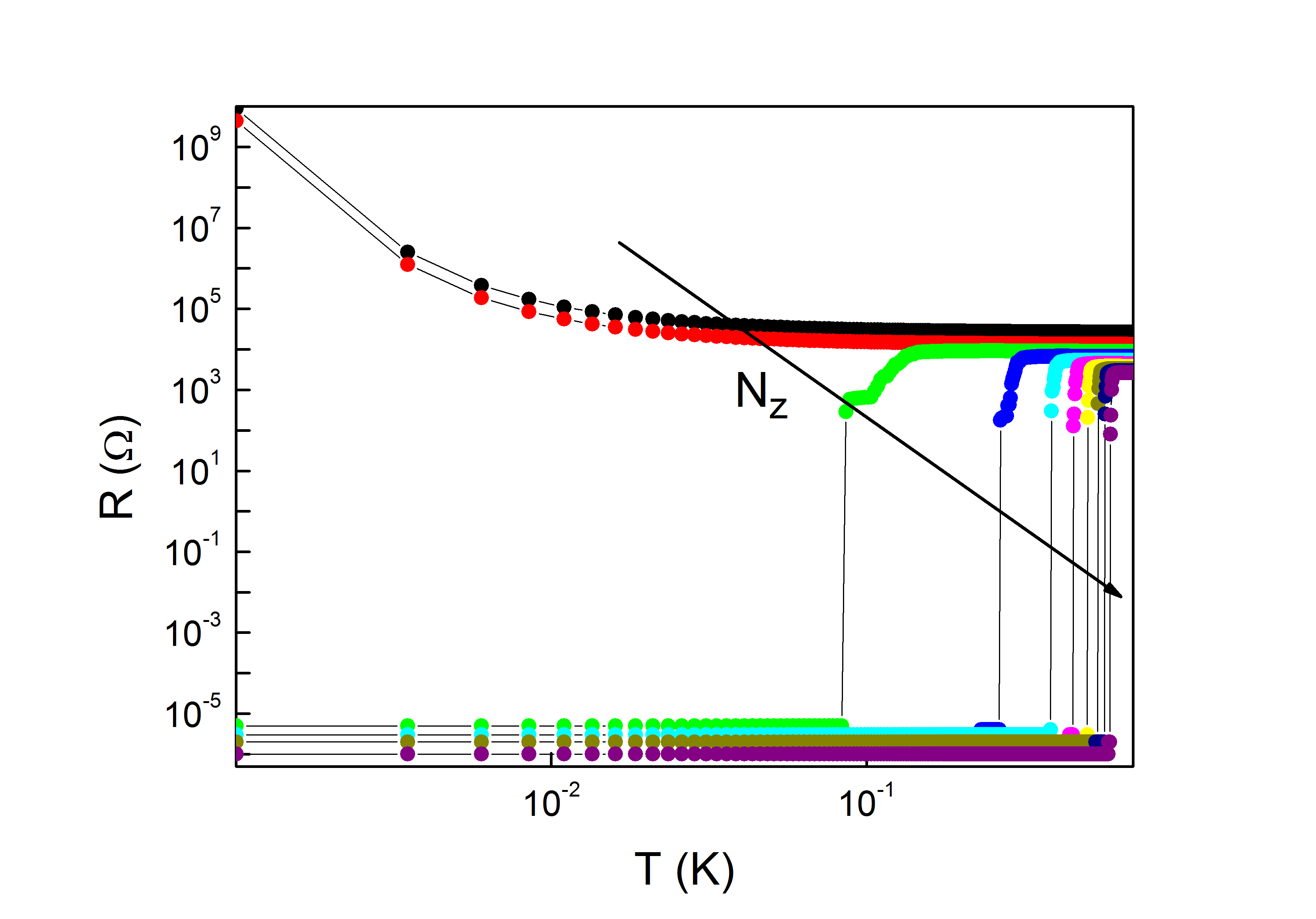}
}
\caption{Superconducting-insulator transition of a three-dimensional network as a function of temperature. For the curves shown in this figure,  $N_{x}=20$, $N_{y}= 25$  and $N_z$ is varied from 1 to 10 in steps of 1 to simulate the film thickness variation. The bias current is kept constant to the value $I=50\mu\mathrm{A}$. The degree of disorder is kept constant with
the standard deviation $\sigma=0.2$.}
\label{PhaseTransitionNzSigma02}
\end{center}
\end{figure}
\begin{figure}
\begin{center}
\resizebox{0.95\columnwidth}{!}{%
\includegraphics{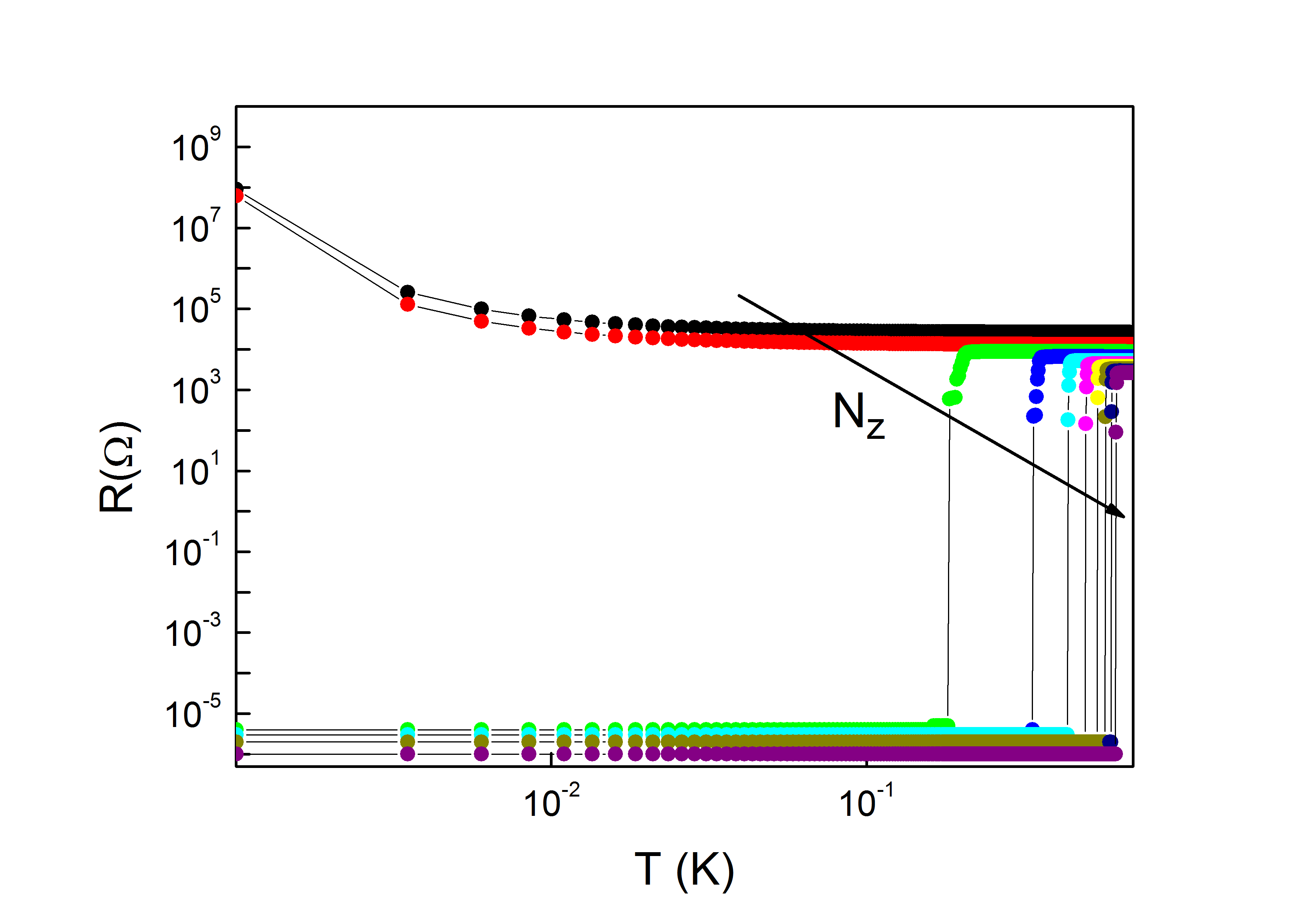}
}
\caption{Superconducting-insulator transition of a three-dimensional network as a function of temperature. For the curves shown in this figure,  $N_{x}=20$, $N_{y}= 25$  and $N_z$ is varied from 1 to 10 in steps of 1 to simulate the  film thickness variation. The bias current is kept constant to the value $I=50\mu\mathrm{A}$. The degree of disorder is kept constant by taking
the standard deviation $\sigma=0.1$.}
\label{PhaseTransitionNzSigma01}
\end{center}
\end{figure}

Figures~\ref{PhaseTransitionNzSigma03},~\ref{PhaseTransitionNzSigma02} and ~\ref{PhaseTransitionNzSigma01}  show the superconductor insulator transition as a function of the temperature for different values of the film thickness, which is varied by changing the number of junctions $N_z$ in the $z$-direction. The bias current is constant ($I=50\mu\mathrm{A}$). In each figure, the degree of disorder is constant,  respectively by $\sigma = 0.3$, $\sigma = 0.2$ and $\sigma = 0.1$.
One can notice that the simulations  are robust with respect to the variation of $N_z$ at different disorder degrees and the insulating phase appears for thinner films (i.e. for thinner networks corresponding to smaller values of $N_z$).
\par
%
% Figura 3

% Figure 4
%

% Figure 5
%
%
% Figure 6
\par
In this paper, the superconducting-insulator transition has been simulated in arrays of thermally activated resistively shunted weak-links at varying disorder, bias current and film thickness. Accurate predictions of the relevant properties of the insulating state  have been obtained.
In particular, the agreement between simulation and experiment is consistent with a complex process according to which the homogeneous system \emph{self-organizes} into  a granular  structure of weak-links \cite{Shahar,Sambandamurthy,Baturina,Baturina2,Sarwa}.
The onset of the insulating phase, whose relevant property is the exponential dependence of the resistance $R$ on temperature $T$, given by  Eq.~(\ref{eq1}),
is recovered,  consistently with the existence of a nonvanishing gap in the insulating phase. Further applications and developments
of this work can be envisioned to shed light on the elementary processes underlying the superconducting-insulator transition. The flexibility and ease of implementation of the proposed approach would be interesting for further investigations of the emergence of such complex electronic structures  in several frameworks \cite{Staley,Salluzzo,Bollinger,Ajmone}.

\end{document}